\documentclass[10pt,twocolumn]{ICCAS2022}
 
\usepackage{diagbox}
\usepackage{cite}

\begin{document}

\title{Construction of the Radio Map with Defective GPS Position Information}

\author{Taewon Kang${}^{1}$ and Joon Hyo Rhee${}^{2*}$ }

\affils{ ${}^{1}$School of Integrated Technology, Yonsei University, \\
Incheon, 21983, Korea (taewon.kang@yonsei.ac.kr) \\
${}^{2}$Korea Research Institute of Standards and Science, \\
Daejeon, 34113, Korea (jh.rhee@kriss.re.kr) \\
{\small${}^{*}$ Corresponding author}}


\abstract{
Over the past few decades, many indoor positioning technologies have been developed. Among them, received signal strength (RSS)-based positioning has attracted attention owing to its simplicity and suitability for various signals such as Long-Term Evolution (LTE), WiFi, and Bluetooth. 
The basic idea of RSS-based indoor positioning is to estimate the receiver location by matching the measured received signal strength indicator (RSSI) with preestablished RSSI collections with corresponding locations, known as the radio map. Hence, constructing an accurate radio map directly relates to accurate positioning performance in RSS-based indoor positioning. 
RSS-based indoor positioning can be easily conducted with a radio map that surveys every location, but a complete radio map cannot be constructed when the map area includes locations that are physically impossible to reach or denied access. 
In addition, measurement errors or device problems can occur during the survey, resulting in degradation of the radio map. We analyzed incidents that occurred in actual RSSI surveys that could disrupt the construction of the radio map and proposed methods to construct accurate radio map.
}


\maketitle


\section{Introduction}
With the increased demand for positioning services, many positioning applications have been developed, including location-based services (LBS) in the use of global navigation satellite systems (GNSS) \cite{Enge11, Lee17:Monitoring, Yoon14:Medium, Lee22:Optimal, Lee22:Nonlinear, Sun21:Markov, Park2021919, Lee22:Urban, Kim2019}, enhanced long-range navigation (eLoran) \cite{Kim22:First, Rhee21:Enhanced, Son20191828, Son2018666, Kim2020796, Park2020824, Hwang2018}, and other sensors \cite{Lee22:SFOL, Kim2017:SFOL, Park2020800, Rhee2019, Kim20181087, Lee2018:Simulation, Rhee2018224, Shin2017617}. However, the performance of many positioning systems is highly degraded in indoor environments, where signal blockage, reflection, or interference frequently occur.

Therefore, many indoor positioning technologies  have been developed over the past few decades \cite{Kang21:Indoor, Kang2020774,Brena2017,Yang2015150,Alarifi2016, Park21:Indoor}. 
Among them, received signal strength (RSS)-based positioning \cite{Lee22:Evaluation} has attracted attention owing to its simplicity and suitability for various signals such as Long-Term Evolution (LTE) \cite{Jia21:Ground, Han2019, Lee2020939, Lee20202347, Lee2020:Preliminary, Jeong2020958, Lee20191187, Kang20191182}, WiFi \cite{Tao201810683}, and Bluetooth \cite{Jianyong2014526}. 
The basic idea of RSS-based indoor positioning is to estimate the receiver location by matching the measured received signal strength indicator (RSSI) with preestablished RSSI collections with corresponding locations, known as the radio map. Hence, constructing an accurate radio map directly relates to accurate positioning performance in RSS-based indoor positioning.

RSS-based indoor positioning can be easily conducted with radio maps that survey every location, but complete radio maps cannot be constructed when the map area includes locations that are physically impossible to reach or denied access, such as private ground. 
In addition, the RSSI must be collected at a precise location at every measurement time to construct an accurate radio map. 
However, measurement errors or device problems can occur during the survey, resulting in degradation of the radio map. 

We analyzed incidents that occurred in actual surveys that could disrupt the construction of the radio map and proposed methods to construct accurate radio map in spite of the incidents.
The analysis and proposed methods are presented in Section 2. The results of the complete radio map construction with the proposed methods are presented in Section 3. Finally, the conclusions of this study are presented in Section 4.

\section{Methodology}

\subsection{Radio map construction using interpolation}

\begin{figure*}
  \centering 
  \includegraphics[width=0.85\linewidth]{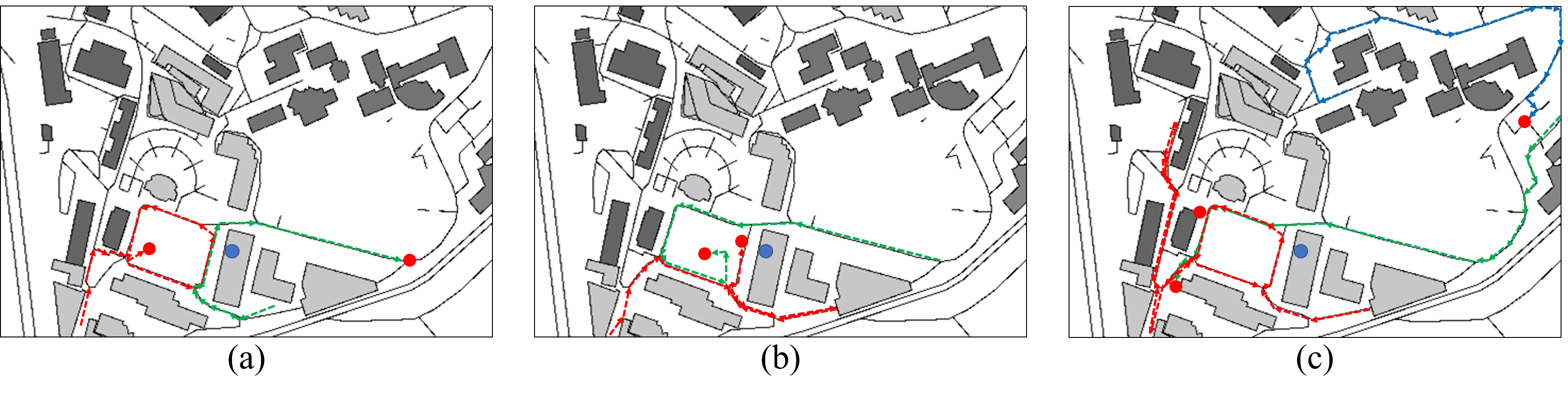}
  \caption{RSSI survey routes of signal strength measurement persons (red dots) for the experiments in (a) Fig. \ref{fig:IDW}, (b) Fig. \ref{fig:pos_error}, and (c) Fig. \ref{fig:no_update_interpolation}. The blue dot indicates the location of a signal transmitter, which remained the same during the three experiments.}
  \label{fig:measure_trajectory}
\end{figure*}

\begin{figure*}
  \centering
  \includegraphics[width=0.85\linewidth]{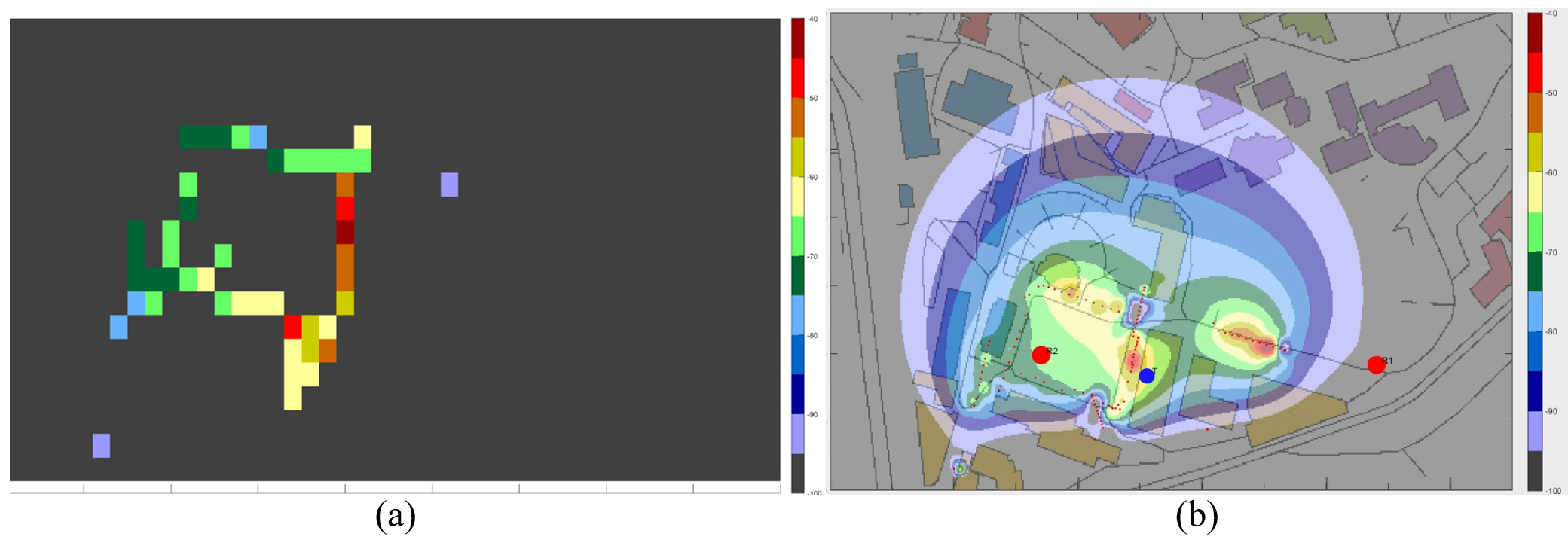}
  \caption{Radio map construction process. (a) Received signal strength indicator (RSSI) value update using EMA. (b) Interpolation using IDW.}
  \label{fig:IDW}
\end{figure*}

To collect the signal strength values used in the radio map, signal strength measurement persons, each carrying a signal receiver and a global positioning system (GPS) receiver, move around a signal transmitter that is placed within the map area. 
The signal receiver measures the RSSI from the transmitted signal. 
For each measurement time $t$, two-dimensional position ${\mathbf{p}}_t=(x_t, y_t)$ provided by the GPS receiver and measured RSSI value $V_t$ are collected.

To construct a radio map, the map area was divided into quadrangular grids of identical shapes and sizes. A basic radio map is equivalent to an array of size that is equal to the number of grids. 
Before the RSSI values were updated in the radio map, radio map grids were initialized by not-a-number (NaN) elements.
The collected RSSI value was then updated in the grid position where the corresponding RSSI was collected. Updates were performed for each measurement time.

When the updates are repeated, a case of updating an RSSI value in the grid that is already filled with a divergent RSSI value can occur. 
If the grid value is overwritten by the up-to-the-minute measured RSSI value, the grid value can be updated to an incorrect value if the RSSI value contains a large measurement error.
Therefore, an exponential moving average (EMA) was applied when updating the RSSI values to a radio map to reduce the effect of temporary measurement errors. 

An updated grid value $S_t$ at $t$-th measurement time is given by (\ref{eqn:EMA}).

\begin{equation}
\label{eqn:EMA}
  S_t=
  \begin{cases}
    V_1, & \text{if}\ t=1 \\
    V_t + (1-\alpha)S_{t-1}, & \text{if}\ t>1
  \end{cases}
\end{equation}
where $\alpha$ is the constant smoothing factor. $\alpha$ is set in a range of $0<\alpha<1$ and increases the influence of up-to-the-minute measured RSSI value $V_t$ and reduces the influence of preexisting grid value $S_{t-1}$ when set close to 1.

When the updates were completed, the radio map consisted of grids filled with the updated RSSI values and NaN elements. Interpolation based on the updated RSSI values was performed to replace the NaN elements by interpolated RSSI values. 
The interpolation method used in this study was inverse distance weighting (IDW) interpolation, specifically the Shepard’s method \cite{Shepard1968517}, that calculates assigned values with a weighted average of the values at known points. Interpolated value $S({\mathbf p})$ at grid point ${\mathbf p}$ is given by (\ref{eqn:IDW}).

\begin{equation}
\label{eqn:IDW} 
  S({\mathbf{p}})=
  \begin{cases}
    \frac{\sum_{i=1}^{N_g} w_i S({{\mathbf{p}}_i})}
         {\sum_{i=1}^{N} w_i}
    = \frac{\sum_{i=1}^{N_g}  \frac{S({{\mathbf{p}}_i})}{d({\mathbf{p}}, {\mathbf{p}}_i)}} 
           {\sum_{i=1}^{N} \frac{1}{d({\mathbf{p}}, {\mathbf{p}}_i)}}, 
    & \text{if}\ d({\mathbf{p}},{\mathbf{p}}_i) \neq 0 \\
    S({{\mathbf{p}}_i}), 
    & \text{if}\ d({\mathbf{p}},{\mathbf{p}}_i)=0 \\
  \end{cases}
\end{equation}
where $N_g$ denotes the total number of grids filled with a measured RSSI value, ${\mathbf p}_i$ denotes a grid point with a non-interpolated RSSI value, and $d({\mathbf p}, {\mathbf p}_i)$ represents the distance between ${\mathbf p}$ and ${\mathbf p}_i$. The weighting factor $w_i$ for IDW is given as $\frac{1}{d({\mathbf p}, {\mathbf p}_i)}$.
After completion of interpolation, a complete radio map is constructed.

\begin{figure*}
  \centering
  \includegraphics[width=0.85\linewidth]{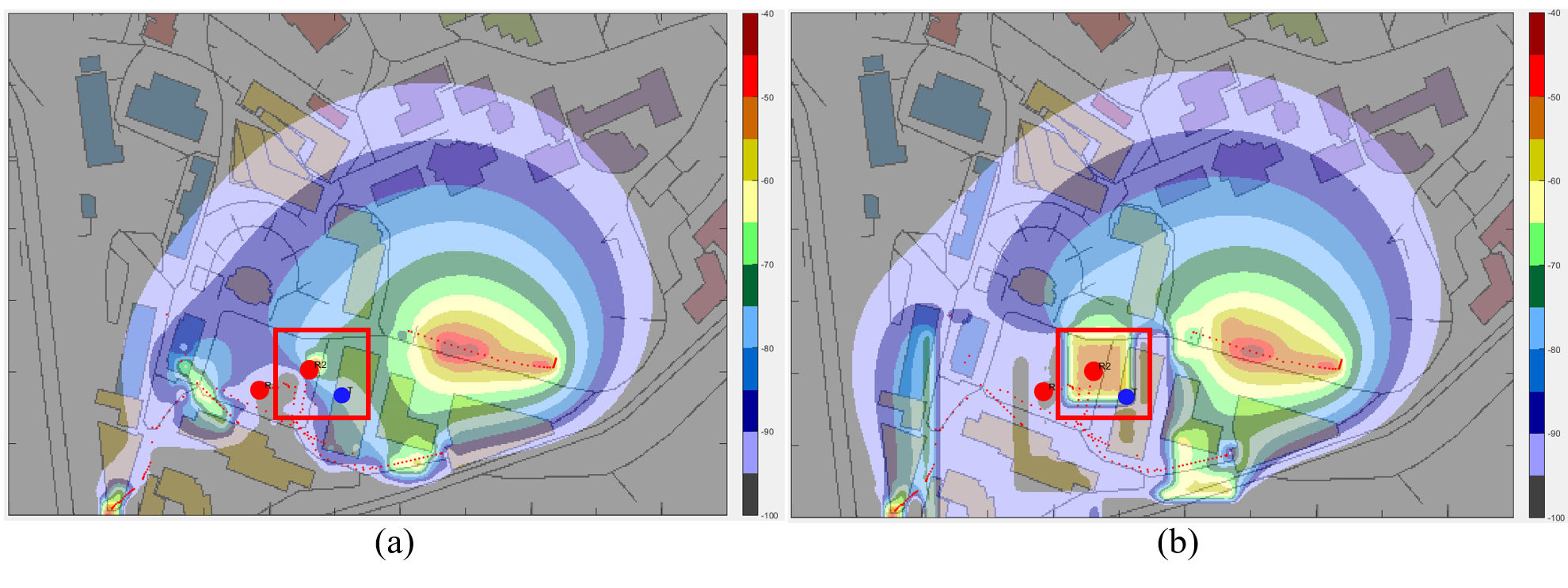}
  \caption{Radio map (a) before considering the receiver position error and (b) after considering the receiver position error.}
  \label{fig:pos_error}
\end{figure*}

\begin{figure*}
  \centering
  \includegraphics[width=0.85\linewidth]{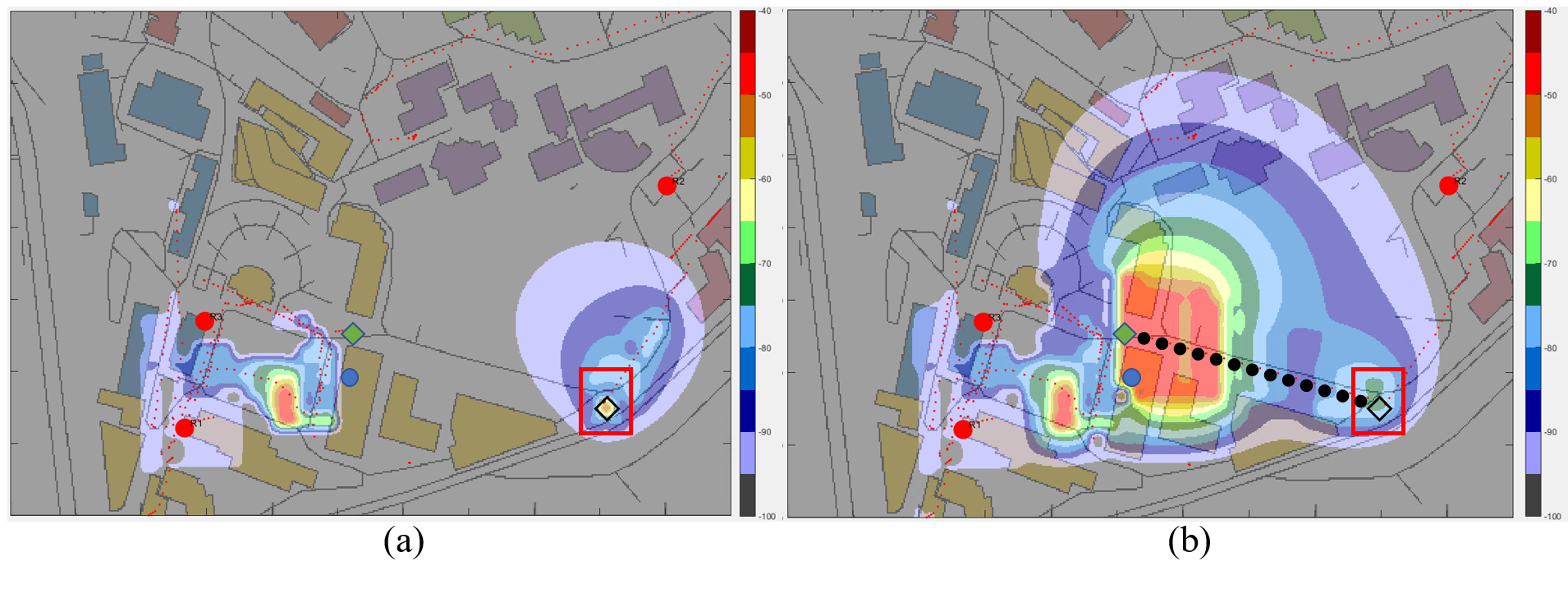}
  \caption{Radio map (a) before position update error correction and (b) after position update error correction.}
  \label{fig:no_update_interpolation}
\end{figure*}

\subsection{Consideration of the receiver position error to the construction of the radio map}
\label{sec:rcvrPosErr}

When the signal strength measurement person moves around an environment that ensures good GPS signal conditions, GPS can provide meter-level accuracy for each RSSI measurement location. 
However, ensuring meter-level positioning accuracy is difficult in environments such as urban canyons or inside buildings, where GPS signals can be reflected or blocked.
This can result in a signal receiver not collecting the accurate position of itself and consequentially constructing an inaccurate radio map. 

To construct a proper radio map under poor GPS positioning accuracy, we applied the standard deviation of GPS position outputs to the process of grid value update in (\ref{eqn:EMA}).
The sample standard deviation of GPS position outputs is shown in (\ref{eqn:std}).

\begin{equation}
\label{eqn:std}
    \sigma=\sqrt{\frac{1}{N_V}\sum_{i=1}^{N_V}(V_t-\mu)^2}
\end{equation}
where $N_V$ denotes the number of previously collected RSSI values that are used for the standard deviation calculation ($N_V = 30$ was used in this study) and $\mu$ denotes the mean value of the RSSI values. 
All grids within a radius of $2 \sigma$ from each GPS position output were updated by each measured RSSI value because the true location of the RSSI measurement could be anywhere within the $2 \sigma$ bound with a high probability.

\subsection{Correction of position update error to the radio map}

Because of a GPS signal reception error, an incident in which the position of the receiver is not updated at the measurement time can occur. 
During this time period, the displayed receiver position remains unchanged although the receiver actually moves.  
In this case, the collected RSSI values from the moving receiver will be continuously reflected in the same grid that corresponds to the unchanged displayed position, which obviously results in the construction of an erroneous radio map.
If the time period of no position update becomes longer, the constructed radio map would become more erroneous.

Assuming that the time period of no position update is fairly short, the radio map can be corrected as soon as the correct receiver position becomes available. 
When the receiver position is updated, a noticeable jump from the previous displayed position to the newly updated position is observed.
We interpolated the points between the newly updated receiver position and the receiver position prior to the update so that the receiver can be depicted as having moved the straight path [e.g., the path with black dots in Fig. \ref{fig:no_update_interpolation}(b)] between the positions with a constant speed. 
The number of interpolated points was set to be the same as the number of measurement epochs at which the position updates were not available. 
Then, each measurement during no position update was assigned to each interpolated point before performing EMA and IDW.

Let $t_s$ denote the time when the position update stopped, $t_f$ denote the time when the position update resumed, and $t_e$ denote the elapsed time since $t_s$. 
Then, the position vector of the interpolated point at time $t_s + t_e$, ${\mathbf{p}}_{t_s + t_e}$, is obtained by (\ref{eqn:p_interp}).

\begin{equation}
\label{eqn:p_interp}
    {\mathbf{p}}_{t_s+t_e} = {\mathbf{p}}_{t_s} + \frac{{\mathbf{p}}_{t_f}-{\mathbf{p}}_{t_s}}{t_f-t_s} t_e
\end{equation}
where $0 \leq t_e \leq t_f-t_s$.


\section{Results}

We conducted RSSI surveys and constructed radio maps of Hanyang University, Seoul, Korea, following the aforementioned radio map construction method. 
The area of interest was 0.7 km $\times$ 1 km. We divided the area into a 10 m $\times$ 10 m square grid to create a 70 $\times$ 100 grid map. 
Two to three persons collected their positions and RSSI values every second along the routes shown in Fig. \ref{fig:measure_trajectory} for each survey case. 
The final positions of signal strength measurement persons after each survey are represented by red dots, and an LTE signal transmitter that was placed inside a building is represented by a blue dot.

Fig. \ref{fig:IDW} shows the process of radio map construction, in order. Fig. \ref{fig:IDW}(a) shows the updated grid map using EMA, and Fig. \ref{fig:IDW}(b) shows the interpolated grid map. Unlike Fig. \ref{fig:IDW}(a), Fig. \ref{fig:IDW}(b) shows a smooth contour line between the different RSSI level areas.

Fig. \ref{fig:pos_error} shows the effect of the position error on the radio map and the corrected result by applying the method explained in Section 2.2. 
The red box area in Fig. \ref{fig:pos_error}(a) contains the location of the signal transmitter (blue dot) and the other area is fairly far from the signal transmitter.
However, a higher RSSI level was observed outside of the red box, which seems to be unnatural. 
This is because the red box area had high GPS positioning errors owing to the nearby buildings that created multipaths.
After considering the GPS positioning errors in the process of grid value update (Section 2.2), the area close to the signal transmitter showed a high RSSI level as intended, as shown in Fig. \ref{fig:pos_error}(b).

Fig. \ref{fig:no_update_interpolation} shows the effect of the position update error on the radio map and the corrected result. The red box in Fig. \ref{fig:no_update_interpolation}(a) shows the area where the position update was delayed due to GPS errors and only the RSSI values were collected and updated. 
As a result, the RSSI level in the red box area became high despite the long distance from the signal transmitter.

After applying the position interpolation technique explained in Section 2.3 between the newly updated receiver position, which is marked by a green diamond, and the receiver position before the update, which is marked by a black edge diamond, a more realistic radio map of Fig. \ref{fig:no_update_interpolation}(b) was obtained.
The high RSSI level area contains the location of the signal transmitter in this case.

\section{Conclusion}

We analyzed incidents that occurred in actual RSSI surveys that could disrupt the construction of a radio map due to the defective GPS position outputs.
In this study, we proposed methods to construct a more accurate radio map under those incidents. 
To demonstrate the effectiveness of the proposed methods, the RSSI surveys and constructions of the radio maps were conducted in a 0.7 km $\times$ 1 km area. 

\section*{ACKNOWLEDGEMENT}

This work was supported by Institute of Information \& Communications Technology Planning \& Evaluation (IITP) grant funded by the Korea government (KNPA) (2019-0-01291, LTE-based accurate positioning technique for emergency rescue).

\bibliographystyle{IEEEtran}
\bibliography{mybibfile, IUS_publications}

\end{document}